# The Impact of Implicit Government Guarantee on Credit Rating of Municipal Investment Bonds

Yan Zhang, Yixiang Tian, Lin Chen

*School of Management and Economics, University of Electronic Science and Technology of China, Chengdu, China*

Corresponding author: chenlin2@uestc.edu.cn

**Abstract**：One type of bond with the most implicit government guarantee is municipal investment bonds. In recent years, there have been an increasing number of downgrades in the credit ratings of municipal bonds, which has led some people to question whether the implicit government guarantee may affect the objectivity of the bond ratings? This paper uses text mining methods to mine relevant policy documents related to municipal investment bond issuance, and calculates the implicit guarantee strength of municipal investment bonds based on the PMC index model. It further analyzes the impact of the implicit guarantee strength of municipal bonds on their credit evaluation. The study found that the implicit government guarantee on municipal investment bonds does indeed help to raise the credit ratings assigned by credit rating agencies. The study found that, moreover, the government's implicit guarantee has a more pronounced effect in boosting credit ratings in less developed western regions.

**Keywords**：PMC Index, Implicit government guarantees, Credit ratings, Municipal investment bonds

## 1.Introduction

The bond market has always been an important sector of the financial market. In recent years, while the bond market in China has been developing rapidly, there have been frequent occurrences of default events in the bond market. Usually, when state-owned enterprises encounter operational difficulties, the government may provide assistance to protect state assets from losses and maintain economic stability through measures such as financial subsidies and capital injections. Although in April 2015, China's first state-owned enterprise bond '11 Tianwei MTN2' failed to repay the principal and interest on time, breaking market expectations of state-owned enterprises being able to fulfill their obligations.However, does the inherent mindset of 'implicit government guarantee' still persist in the market?And whether this 'implicit government guarantee' affects the pricing and default risk of state-owned enterprises is something we must pay attention to.

Relevant studies have shown that implicit guarantees alter investors' perceptions and may lead to economic distortions.In essence, the rigid redemption belief is a disregard for risk, and the default of state-owned enterprise bonds to some extent benefits the return of risk awareness and forms a more perfect market mechanism.However, the significant role of state-owned enterprises in the development of China's national economy can easily lead to cross-contagion of default risks and trigger systemic financial risks. The impact brought by concentrated defaults on bonds issued by state-owned enterprises should not be ignored.For example, since the second half of 2020, the number of defaulted bonds issued by local and central state-owned enterprises has exceeded one-third of the total defaults, resulting in a significant drop in prices for a large number of high-grade credit bonds. This has not only impacted market confidence but also led to the cancellation or postponement of credit bond issuances worth hundreds of billions.

At present, there is no consensus on the impact of implicit government guarantees on the default of state-owned enterprise bonds.Some researchers believe that the implicit guarantee provided by local governments enhances the financing capability of state-owned enterprises, benefiting their operations and thus reducing default risks.Other researchers, however, argue that the government is not necessarily the main entity responsible for debt repayment. Implicit government guarantees may not necessarily reduce credit risk for businesses and their impact on credit spreads varies across regions.Although implicit guarantees have to some extent improved the financing capacity of state-owned enterprises, in the long run, government implicit guarantees have led to market participants gradually forming investment expectations of rigid repayment for state-owned enterprise bonds.While increasing investment risks, it also leads to the flow of social funds towards enterprises with excess production capacity, which may ultimately result in a mismatch of social resources and the risk of corporate debt default.

The most government implicitly guaranteed type of bonds are municipal investment bonds, also known as 'quasi-municipal bonds'. They are issued by local financing platforms as the issuing entities and publicly issue corporate bonds and medium-term notes, mainly used for local infrastructure construction or public welfare projects. Therefore, local governments usually take a series of measures to enhance the repayment capacity of enterprises or ensure the source of debt repayment funds in order to reduce issuer credit risk. So, local government credit provides implicit guarantee for municipal investment bonds. In the past decade, the scale of issuance for municipal investment bonds has grown rapidly, with annual issuances increasing from 358.6 billion yuan in 2009 to 4.89 trillion yuan in 2023. While providing significant financial support for local infrastructure investments, this has also raised concerns about the risk of local government debt due to implicit guarantees.

Unlike explicit guarantee methods such as joint liability guarantees, shortfall compensation, mortgage guarantees, and pledge guarantees, government implicit guarantees typically manifest as the local government's decision to provide financial subsidies or debt swaps to repay local investment bonds in order to mitigate risks. This type of guarantee is not legally binding and the decision on whether or how much assistance is provided rests solely with the local government based on its own interests. However, when companies face credit crises or significant financial risks, governments often provide financial assistance to prevent systemic risks and stabilize the economy (Luo Ronghua and Liu Jinjin, 2016; Ma Wentao and Ma Caoyuan [2], 2018; Zhang Lu, 2020) [3-4].Therefore, the existence of implicit government guarantees may lead to investors blindly underestimating the risks of city investment platform companies, resulting in a deviation between the pricing of city investment bonds and their fundamentals (Zhong Ninghua, Chen Shanshan, Ma Huixian et al.[5] ,2021),which in turn reduces the cost of issuing bonds for city investment platforms (Luo Ronghua and Liu Jinjin, 2016), promotes rapid expansion of local government's implicit debt scale (Ma Wentao and Ma Caoyuan, 2018), and brings uncertainty and risk to financial markets(Mao Rui, Liu Nannan, Liu Rong et al.[6]，2018）.From this, it can be seen that conducting a thorough analysis of the impact of implicit government guarantees on local government bonds is helpful in promoting the withdrawal of implicit government guarantees from the bond market and is of great significance in preventing and resolving risks associated with local government debt.In summary, existing research has shown that implicit government guarantees have a significant impact on the issuance yield of local government bonds (Wang Li and Chen Shiyi, 2015; Luo Ronghua and Liu Jinjin, 2016), secondary market risk premium (Zhong Ninghua, Chen Shanshan, Ma Huixian et al., 2021; Zhu Xiaoquan, Chen Zhuo and Shi Zhan, 2022), default risk (Wang Li and Chen Shiyi[7], 2015; Yuan Leping and Xiao Yan[8], 2017; Zhu Xiaoquan, Chen Zhuo and Shi Zhan[9], 2022), and credit ratings (Zhong Huiyong, Zhong Ninghua and Zhu Xiaoneng[10], 2016).

The manifestation of credit risk is not only limited to default events, but also primarily reflected in the changes of credit ratings in the bond market.In recent years, with the continuous default of state-owned enterprise bonds with high credit ratings, there has been an increasing frequency of credit rating downgrades for municipal investment bonds.Seemingly raising doubts, does the implicit government guarantee affect the objectivity of city investment bond ratings?The study conducted by Zhong Huiyong et al. (2016) indicates that "implicit guarantees" are beneficial for improving the credit ratings of bonds, and implicit guarantees have different impacts on local government financing vehicle bonds with different credit ratings [11] [12].Therefore, in the context of increasing downgrades in credit ratings for local government financing vehicle bonds and the continuous accumulation of credit risks, it is of vital importance to explore the impact mechanism of implicit government guarantees on the credit ratings of these bonds in order to understand the sources and mitigation measures for credit risks associated with local government financing vehicle bonds.This will provide reference for the government to further regulate the municipal investment bond market, help municipal investment companies improve their own management, enable investors to have a deeper understanding of municipal investment risks, and encourage all stakeholders to work together to prevent the risk of credit rating downgrades for municipal investment bonds and promote the healthy development of the municipal investment bond market.In conclusion, the objectivity and fairness of credit ratings for municipal investment bonds are of profound significance to the development of the rating industry itself as well as the construction of the entire social credit system.

This article constructs a PMC index model for implicit guarantee of municipal investment bonds and explores the impact of implicit guarantees on credit ratings of municipal investment bonds using a panel ordered Logit model. The second section provides a brief introduction to the panel ordered logit model. The third section introduces the calculation of the government's implicit guarantee intensity using the PMC index model. The fourth section conducts an empirical study on the credit rating of local government financing vehicle bonds affected by implicit guarantees. Finally, the conclusion summarizes the entire article.

## 2.Panel ordered Logit model

### 2.1 basic model

Credit rating is a typical ordinal qualitative variable, and even in large samples, linear models tend to have biases when estimating credit ratings. In contrast, discrete choice models can directly use the original information of credit ratings without the need for linear transformation that may cause distortion and bias.In the case where the dependent variable is a categorical variable, if it is a binary choice variable, estimation should be conducted using the logit model. However, if the dependent variable is a multicategory variable with an ordered structure, an ordered logit model (OLM) should be constructed for analysis.The ordered Logit model is a regression model for ordinal response variables, which is based on the cumulative probability function of the response variable. It assumes that the logit function of each cumulative probability is a linear function of covariates, with constant regression coefficients.

Let $y_i$ be the ordered response variable of the $i$ th object over C classes and a series of covariables $x_i$.The mapping between covariable $x_i$ and class probability $p_{ci}$ is defined as follows:

$$p_{ci} = Pr(y_i = y_c|x_i), \quad c = 1, 2, \ldots, C$$

The ordered Logit model is not represented by categorical probabilities, but by a cumulative probability $g_{ci}$ that facilitates one-to-one transformations:

$$g_{ci} = Pr(y_i \leq y_c|x_i), \quad c = 1, 2, \ldots, C$$

Since the final cumulative probability is 1, in this case the model specifies only the cumulative probability of C-1. An ordered response variable $Y_i$ of class C is defined as a C-1 equation, and the cumulative probability $g_{ci} = Pr(y_i \leq y_c / x_i)$ is related to the results of linear prediction. An ordered Logit model is constructed, as shown in equation (1):

$$logit(g_{ci}) = log(\frac{g_{ci}}{1-g_{ci}}) = \alpha_c - x_i'\beta, i = 1,2,\ldots,n \tag{1}$$

Parameter $\alpha_c$ is the cutoff point, and the covariable $\beta$ used to represent the addition of a positive slope is related to the movement of the response variable, further indicating that the probability of moving to a higher category increases. The cumulative probability of class C is shown in equation (2):

$$g_{ci} = \frac{\alpha_c - x_i'\beta}{1+e^{\alpha_c - x_i'\beta}} = \frac{1}{1+e^{-\alpha_c + x_i'\beta}} \tag{2}$$

A sequence C-1 unobservable potential variable $y_i^*$ will be generated by classifying the ordered response variable $y_i$ by C, then an ordered Logit model is constructed as shown in equation (3):

$$y_i^* = x_i'\beta + \varepsilon_i, i = 1,2,\ldots,n \tag{3}$$

The relationship between credit rating, which is the observable variable $y_i$, and the latent variable $y_i^*$ is shown in equation (4):

$$y_i = \begin{cases} 1 & if & y_i^* \in \tau_1 \equiv (-\infty, a_1] \\ 2 & if & y_i^* \in \tau_2 \equiv (a_1, a_2] \\ \vdots & \vdots & \vdots \\ c & if & y_i^* \in \tau_c \equiv (a_{c-1}, a_c] \\ \vdots & \vdots & \vdots \\ C & if & y_i^* \in \tau_C \equiv (a_{C-1}, +\infty) \end{cases}, c = 1,2,\ldots,C \tag{4}$$

Where, the number of truncation points $a_1, a_2, \cdots, a_{C-1}$ estimated by the model is determined by the number of categories of observable variable $y_i$, and the truncation points are arranged in increasing order, that is, $\alpha_1 < \alpha_2 < \ldots < \alpha_{C-1}$. Suppose $s$ is the state matrix of $y_i$, then, $s = [s_1, s_2, \cdots, s_c, \cdots, s_C]' = [1, 2, \cdots c, \cdots, C]'$, then the probability of $y_i$ under state $s$ is shown in formula (5):

$$Pr(y_i = s_c | x_i) = Pr(y_i^* \in \tau_c | x_i) = Pr(x_i'\beta + \varepsilon_i \in \tau_c | x_i), c = 1,2,\ldots,C \tag{5}$$

If the error term $\varepsilon_i$ follows the logistic function, then the model is an ordered Logit model. Suppose $F(\cdot)$ is the cumulative distribution function of $\varepsilon_i$, the function is shown in equation (6), $C - 1$ truncation points $a_1, a_2, \cdots, a_{C-1}$ divide the probability density function into c intervals:

$$F(\varepsilon_i | x_i) = F(\varepsilon_i) = \frac{e^\varepsilon}{1+e^\varepsilon} = \Phi(\varepsilon_i) \tag{6}$$

Further rewrite equation (5) as equation (7):

$Pr(y_i = s_c | x_i) =$

$$\begin{cases} \Phi\left(\frac{\alpha_1-x_i'\beta}{\sigma}\right) & if \quad c=1 \\ \vdots & \vdots \quad \vdots \\ F(\alpha_c - x_i'\beta|x_i) - F(\alpha_{c-1} - x_i'\beta|x_i) = \Phi\left(\frac{\alpha_c-x_i'\beta}{\sigma}\right) - \Phi\left(\frac{\alpha_{c-1}-x_i'\beta}{\sigma}\right) & if \quad 1<c<C \\ \vdots & \vdots \quad \vdots \\ 1 - \Phi\left(\frac{\alpha_{c-1}-x_i'\beta}{\sigma}\right) & if \quad c=C-1 \end{cases} \tag{7}$$

The estimates of parameters $\beta$ and $C-1$ thresholds are derived from the maximized log-likelihood function. In formula (7), since the same set of $\beta$ is used, the probability of $y_i = s_c$ is constant under the influence of $x_i$, that is, the parallel line hypothesis. Under the premise of this hypothesis, the ordered Logit model can be effectively estimated.

## 2.2 Panel order model of municipal investment bonds credit rating

Taking the credit rating of municipal investment bonds as the observable variable $y_i$, and incorporating the unobservable potential variable $y_i^*$ into the model, this paper builds an ordered Logit model as shown in equation (8):

$$y_i^* = x_i'\beta + \varepsilon_i, i = 1,2,\ldots,n \tag{8}$$

Due to the government's requirement for credit ratings of local government financing vehicle bonds, which must be at least AA rating or above, there are actually only three credit ratings for such bonds: AAA, AA+, and AA. Let the credit rating be the observable variable $y_i$, and the relationship between the latent variable $y_i^*$ is shown in equation (9):

$$y_i = \begin{cases} AAA & if \quad y_i^* \leq a_1 \\ AA^+ & if \quad a_1 < y_i^* \leq a_2 \\ AA & if \quad y_i^* > a_2 \end{cases} \tag{9}$$

When assigning credit ratings to the bonds issued by municipal investment companies, they are ranked from highest to lowest, with AAA being assigned a value of 1 and AA being assigned a value of 3. Therefore, equation (9) can be further transformed into equation (10):

$$y_i = \begin{cases} 1 & if \quad y_i^* \leq a_1 \\ 2 & if \quad a_1 < y_i^* \leq a_2 \\ 3 & if \quad y_i^* > a_2 \end{cases} \tag{10}$$

If $s$ is the state matrix of $y_i$, $s = [s_1, s_2, s_3]' = [1,2,3]'$, then the probability of $y_i$ in state $s$ is shown in the formula (4-11):

$$\begin{aligned} &Pr(y_i = s_c|x_i) \\ &= \begin{cases} Pr(y_i^* \leq a_1|x_i) = Pr(x_i'\beta + \varepsilon_i \leq a_1|x_i) = Pr(\varepsilon_i \leq a_1 - x_i'\beta|x_i) \\ Pr(a_1 < y_i^* \leq a_2|x_i) = Pr(a_1 < x_i'\beta + \varepsilon_i \leq a_2|x_i) \\ Pr(y_i^* > a_2|x_i) = Pr(x_i'\beta + \varepsilon_i > a_2|x_i) = Pr(\varepsilon_i > a_2 - x_i'\beta|x_i) \end{cases} \end{aligned} \tag{11}$$

If the error term $\varepsilon_i \sim N(0, \sigma)$, then equation (11) is rewritten as equation (12):

$$Pr(y_i = s_c | x_i) = \begin{cases} \Phi(\frac{\alpha_1 - x_i'\beta}{\sigma}) \\ \Phi(\frac{\alpha_2 - x_i'\beta}{\sigma}) - \Phi(\frac{\alpha_1 - x_i'\beta}{\sigma}) \\ 1 - \Phi(\frac{\alpha_2 - x_i'\beta}{\sigma}) \end{cases} \quad (12)$$

## 3.Implicit guarantee measure based on PMC index

The Policy Modeling Consistency (PMC) index model can obtain raw data through text mining. Due to the absence of restrictions on variable selection and weighting, it can largely avoid subjectivity and has advantages such as high evaluation accuracy and strong operability. Therefore, this model has become a popular tool for evaluating policy effectiveness.In the financial market, all bond products are interconnected with other market factors. Therefore, by analyzing the main policy documents related to local government financing vehicle bonds and using text analysis methods, we can uncover the PMC index model that reflects the implicit guarantee willingness and strength of the government in supporting these bonds.The PMC index model is a mainstream policy evaluation tool in the pre-financial market.The PMC index model for constructing the government implicit guarantee policy of municipal investment bonds mainly includes four steps: first, variable selection and parameter identification; second, establishment of a multi-input-output table; third, calculation of the PMC index; fourth, calculation of the government's implicit guarantee strength.

In this study, in order to quantitatively measure the implicit guarantee policy of government for municipal investment bonds, we first extensively collected relevant policy documents on municipal investment bonds issued by national-level departments such as the State Council, Ministry of Finance, People's Bank of China, and National Development and Reform Commission.Since the establishment of the municipal investment bond market, a total of 17 policy documents have been carefully selected and compiled as analysis samples.Utilize a text content mining system to perform word segmentation on these policy texts, and calculate the word frequency. Then, filter out high-frequency vocabulary and keywords that are closely related to the research topic.Based on the existing PMC index model, we have established an evaluation system for government implicit guarantee policies of municipal investment bonds.

The PMC index evaluation system has established the following 10 primary variables: nature of policy (P1), timeliness of policy (P2), policy area (P3), issuing institution (P4), incentive guarantee (P5), policy function (P6), level of impact (P7), target object of policy (P8), effectiveness level (P9) and policy transparency(P10).Specifically, P1, P2, P4, P9, and P10 assess the quality and transparency of policies from the perspective of their basic attributes. On the other hand, P3, P5, P6, P7, and P8 identify the ability of policies to address and regulate implicit guarantee issues in the local government bond market based on their coverage in different areas, incentive mechanisms, functional positioning as well as their impact range and target audience.In the primary variable, further subdivision of corresponding secondary variables resulted in a total of 47 secondary variables (see Appendix Table 1).After determining the first and second-level variables, the calculation of the PMC index for government implicit guarantee policy of local government financing bonds is as follows:

$$PMC = \sum_{i=1}^{10} \sum_{j=1}^{n_i} \frac{P_{ij}}{n_i} \qquad (1)$$

$P_{ij}$ represents the value of the J th second-level indicator among the i first-level indicators, and $n_i$ represents the number of second-level indicators. When the policy document meets the requirements of the corresponding variables, $P_{ij}$ variable is assigned a value of 1. If not, $P_{ij}$ is assigned to 0.

On the basis of formula (1), the government implicit guarantee strength index $G$ is calculated in formula (2).The series of policies introduced by various levels of government and departments regarding the management of local debt and municipal investment bonds aim to eliminate investors' expectations of implicit government guarantees and reduce the risks associated with municipal investment bonds.Therefore, if these policies have a good effect in reducing the expected effect of implicit guarantee and the risk of municipal investment bonds is reduced, the lower the potential demand for implicit government guarantee will be, and the intensity of implicit government guarantee will decrease accordingly.Therefore, based on the PMC index of municipal investment bonds policy and the distribution of PMC index, the government implicit guarantee strength index $G$ is defined as follows:

$$G = 10 - PMC \qquad (2)$$

Finally, we obtained the government implicit guarantee strength index G of the municipal investment bond PMC index between 2008 and 2024 as shown in Figure 1.

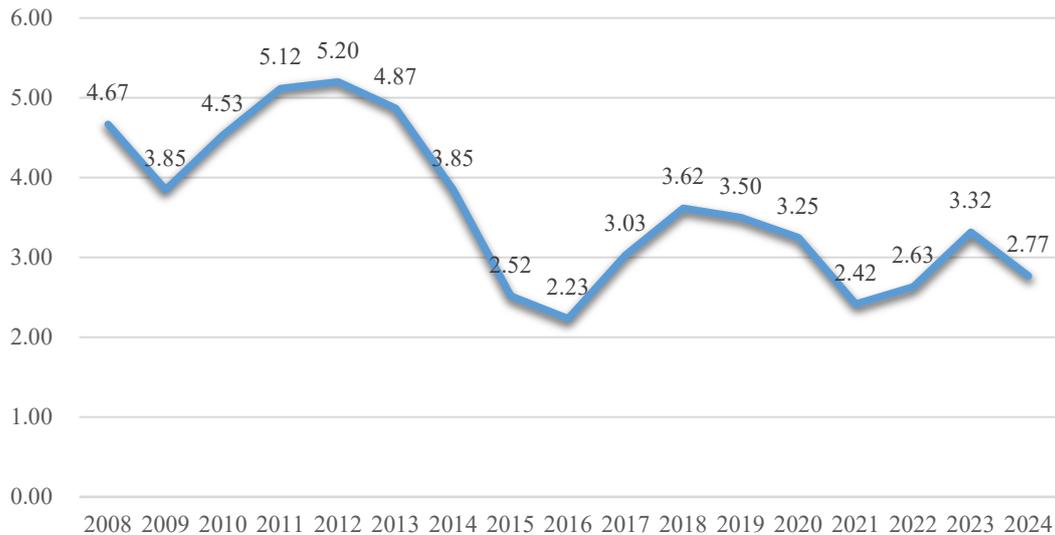

**Figure 1 Variation trend of government implicit guarantee strength index**

In general, the index of implicit guarantee intensity fluctuated from 4.67 points in 2008 to 2.77 points in 2024.Before 2013, the intensity index was relatively high, indicating that during this period the government had a significant implicit guarantee for local government bonds.After 2014, the index tended to stabilize and remained at a low level, which was attributed to policy adjustments, enhanced market self-regulation capabilities, and the government's reassessment of risk control.In the next section, we will analyze whether this index affects the credit rating of municipal bond companies.

## 4. Empirical study on the impact of implicit guarantee on the rating of municipal investment bonds

### 4.1 data specification

This article selected a sample data of 36,472 sets of municipal investment bonds issued by all municipal investment companies in the public bond market from January 1, 2015 to December 31, 2023. After removing outliers and missing values, a total of 9,788 sets of sample data were obtained. The credit ratings of municipal investment bonds were evaluated by credit rating agencies.The final selection of variables is summarized in Table 1.

Table 1 Summary of variables

| variable type | variable | variable declaration |
|---|---|---|
| explained variable | i_ra | The bond rating of the Municipa Investment bonds at the time of issuance will be assigned from high to low, the highest is AAA, assigned 1, the lowest is AA, assigned 3 |
| explaining variable | im_gurantee | Government implicit guarantee strength based on PMC index model G |
| control variable | amount | Total issuance, unit: 100 million yuan |
| Municipal investment bond characteristic variable control variable | term | Issue period, unit: year |
| | Option | Whether it contains option, if yes, take 1; otherwise, take 0 |
| | ROA | Profitability, return on total assets of the issuer, unit: % |
| Platform financial variable control variable | DTA | Solvency, debt-asset ratio of the issuer, unit: % |
| | AT | Operating capacity, total asset turnover of the issuer, unit: times |
| macroeconomic variable | GDP_growth | GDP growth rate last year, unit: % |

The descriptive statistics of the variables are shown in Table 2.

Table2 Descriptive statistics of major variables

| variable | N | Mean | Std.dev. | Min | Max |
|---|---|---|---|---|---|
| i_ra | 9,788 | 1.9376 | 0.7976 | 1.0000 | 3.0000 |
| im_gurantee | 9,788 | 0.2919 | 0.0476 | 0.2233 | 0.3617 |
| amount | 9,788 | 8.4809 | 5.1610 | 0.2800 | 72.0000 |
| term | 9,788 | 5.7734 | 1.9760 | 1.0000 | 24.0000 |
| option | 9,788 | 0.7453 | 0.4357 | 0.0000 | 1.0000 |
| ROA | 9,788 | 1.5996 | 1.1095 | -6.0477 | 14.0272 |
| DTA | 9,788 | 52.8133 | 13.7823 | 2.8412 | 88.6412 |
| AT | 9,788 | 0.0713 | 0.0639 | 0.0002 | 1.7553 |
| GDP_growth | 9,788 | 0.0802 | 0.0335 | 0.0274 | 0.1339 |

The correlation between variables is shown in Table 3. The correlation is not strong and the possibility of mutual interference between variables is small.

Table 3 correlative coefficient matrix

| variable | im_gurantee | amount | term | option | ROA | DTA | AT | GDP_growth |
|---|---|---|---|---|---|---|---|---|
| im_gurantee | 1.0000 | | | | | | | |
| amount | -0.1007 | 1.0000 | | | | | | |
| term | -0.0920 | 0.1093 | 1.0000 | | | | | |
| option | -0.0903 | -0.0165 | 0.5234 | 1.0000 | | | | |
| ROA | -0.0912 | 0.0053 | 0.0865 | 0.0935 | 1.0000 | | | |
| DTA | 0.1006 | 0.0405 | -0.2735 | -0.2395 | -0.2763 | 1.0000 | | |
| AT | -0.0399 | -0.0540 | -0.0470 | -0.0003 | 0.3591 | 0.0342 | 1.0000 | |
| GDP_growth | 0.3604 | -0.0256 | 0.0507 | -0.0035 | 0.0462 | -0.0412 | -0.0133 | 1.0000 |

As shown in Table 4, the distribution of ratings for each year is different. From 2015 to 2023, the number of AAA-rated bonds increased from 50 to 512, indicating a rising trend and demonstrating an annual increase in the issuance of high-rated local government financing vehicles. The number of AA+ rated local government financing vehicle bonds has remained relatively stable, increasing from 162 to 204, with overall minor changes. The number of AA-rated municipal investment bonds has been decreasing year by year, dropping from 419 to 25, indicating a significant reduction in the issuance of low-rated municipal investment bonds.

Table 4 Municipal Investment Bond Rating Distribution (2015-2023)

| Year | 2015 | 2016 | 2017 | 2018 | 2019 | 2020 | 2021 | 2022 | 2023 | Total |
|---|---|---|---|---|---|---|---|---|---|---|
| AAA | 50 | 165 | 216 | 290 | 357 | 513 | 690 | 645 | 512 | 3,438 |
| AA+ | 162 | 312 | 300 | 283 | 418 | 594 | 760 | 490 | 204 | 3,523 |
| AA | 419 | 564 | 457 | 299 | 389 | 337 | 224 | 113 | 25 | 2,827 |
| Total | 631 | 1,041 | 973 | 872 | 1,164 | 1,444 | 1,674 | 1,248 | 741 | 9,788 |

## 4.2 empirical result

As shown in Table 5, the complete estimation results of constructing an ordered Logit model are as follows. The signs of all parameters for the eight variables are consistent with expectations, and the regression results are significant. The overall model is also significant, indicating that the model has a certain explanatory power.

The results of explaining the variable government implicit guarantee strength (im_gurantee)

show that the implicit guarantee strength of the government has a significant negative impact on bond rating values, and therefore has a significant positive impact on bond ratings. That is, the greater the government's guarantee strength, the lower the rating value and correspondingly higher rating level.

The variables of total issuance amount, term, and debt servicing ability have a significant negative impact on the credit rating value. The larger the factor values are, the smaller the rating value becomes, indicating a higher rating. The presence of options, profitability (ROA), operational capability (AT), and GDP growth rate (GDP_growth) all show a significant positive impact on the rating value. The larger the factor values, the higher the bond rating value and the lower the rating.

Table 5 OLM model estimation results

| i_ra | coefficient | Std.dev | t-value | p value | 95% confidence interval | |
|---|---|---|---|---|---|---|
| im_gurantee | -3.781*** | 0.439 | -8.610 | 0.000 | -4.641 | -2.920 |
| amount | -0.051*** | 0.004 | -12.80 | 0.000 | -0.059 | -0.043 |
| term | -0.150*** | .0127 | -11.85 | 0.000 | -0.175 | -0.125 |
| option | 0.115** | 0.052 | 2.220 | 0.026 | 0.013 | 0.217 |
| ROA | 0.096*** | 0.019 | 5.040 | 0.000 | 0.058 | 0.133 |
| DTA | -0.030*** | 0.002 | -19.03 | 0.000 | -0.033 | -0.027 |
| AT | 0.586* | 0.309 | 1.900 | 0.058 | -0.019 | 1.191 |
| GDP_growth | 4.383*** | 0.609 | 7.200 | 0.000 | 3.190 | 5.577 |
| cut1 | -4.007 | 0.181 | | | -4.361 | -3.653 |
| cut2 | -2.379 | 0.178 | | | -2.727 | -2.031 |

## 4.3 robustness examination

In order to further test the robustness of the model, this study replaces the ordered Logit model with a multinomial Logit model for estimation, as shown in Table 6. Compared to the ordered Logit model, the overall pseudo R2 of this model increases slightly from 0.042 to 0.045, indicating a slight improvement. However, the significance of individual variables decreases to varying degrees, such as return on assets (ROA) changing from significant to insignificant and some variable coefficients changing signs. Therefore, in general, the ordered Logit model constructed in this paper is robust when exploring the impact of government implicit guarantee on the credit rating of municipal investment bonds.

Table 6 Comparison of estimation results of multi-class logit model

| i_ra | ordinal number | coefficient | Std.dev. | significance | ordinal number | coefficient | Std.dev | significance |
|---|---|---|---|---|---|---|---|---|
| im_gurantee | 1 | 3.435 | 0.565 | *** | 3 | -1.498 | 0.588 | ** |
| amount | 1 | 0.039 | 0.005 | *** | 3 | -0.034 | 0.006 | *** |
| term | 1 | 0.145 | 0.015 | *** | 3 | -0.056 | 0.019 | *** |
| option | 1 | -0.192 | 0.066 | *** | 3 | -0.057 | 0.073 | |
| ROA | 1 | -0.023 | 0.027 | | 3 | 0.117 | 0.026 | *** |
| DTA | 1 | 0.014 | 0.002 | *** | 3 | -0.026 | 0.002 | *** |
| AT | 1 | -0.134 | 0.427 | | 3 | 0.661 | 0.426 | |
| GDP_growth | 1 | -1.494 | 0.791 | * | 3 | 4.659 | 0.822 | *** |
| Constant | 1 | -2.689 | 0.233 | *** | Constant | 1.564 | 0.240 | *** |
| baseline group | 2 | | | | | | | |
| Pseudo R-squared | | | | | | 0.045 | | |

## 4.5 Analysis of heterogeneity

Due to the varying levels of development and government financial conditions between cities in the eastern region and those in the central and western regions, their implicit guarantee capabilities differ. In addition, differences in financing difficulty for issuers of municipal investment bonds and significant variations in liquidity may result in different repayment pressures for these bonds, leading to differences in default risk. Therefore, in order to explore the different impacts of implicit government guarantees on local municipal investment bond credit ratings in different regions, this section divides the regions into eastern and central-western regions. Using the same method, empirical analyses are conducted separately for these two regions, with the results shown in Table 7 and Table 8 respectively.

Table 7 Regression results of the eastern region

| i_ra | coefficient | Std.dev. | t-value | p value | 95% conf. interval | |
|---|---|---|---|---|---|---|
| im_gurantee | -2.9386*** | 0.6527 | -4.50 | 0.000 | -4.2180 | -1.6593 |
| amount | -0.0893*** | 0.0062 | -14.52 | 0.000 | -0.1014 | -0.0773 |
| term | -0.1339*** | 0.0180 | -7.46 | 0.000 | -0.1690 | -0.0987 |
| option | 0.1539** | 0.0713 | 2.16 | 0.031 | 0.0142 | 0.2936 |
| ROA | 0.1614*** | 0.0283 | 5.70 | 0.000 | 0.1059 | 0.2170 |
| DTA | -0.0255*** | 0.0025 | -10.16 | 0.000 | -0.0305 | -0.0206 |
| AT | 0.1854 | 0.4516 | 0.41 | 0.681 | -0.6997 | 1.0704 |
| GDP_growth | 3.9984*** | 0.8926 | 4.48 | 0.000 | 2.2490 | 5.7479 |
| cut1 | -3.6787 | 0.2737 | | | -4.2151 | -3.1423 |
| cut2 | -1.8177 | 0.2693 | | | -2.3456 | -1.2898 |

Table 8 Regression results of the central and western regions

| i_ra | coefficient | Std.dev. | t-value | p value | 95% conf. interval | |
|---|---|---|---|---|---|---|
| im_gurantee | -3.9064*** | 0.6087 | -6.42 | 0.000 | -5.0995 | -2.7133 |
| amount | -0.0154*** | 0.0054 | -2.83 | 0.005 | -0.0261 | -0.0047 |
| term | -0.1765*** | 0.0184 | -9.59 | 0.000 | -0.2125 | -0.1404 |
| option | -0.0324 | 0.0795 | -0.41 | 0.683 | -0.1882 | 0.1233 |
| ROA | 0.0452* | 0.0266 | 1.70 | 0.089 | -0.0069 | 0.0973 |
| DTA | -0.0295*** | 0.0021 | -13.92 | 0.000 | -0.0337 | -0.0253 |
| AT | 0.6341 | 0.4381 | 1.45 | 0.148 | -0.2246 | 1.4928 |
| GDP_growth | 4.6127*** | 0.8433 | 5.47 | 0.000 | 2.9600 | 6.2655 |
| cut1 | -4.0658 | 0.2500 | | | -4.5557 | -3.5759 |
| cut2 | -2.5865 | 0.2459 | | | -3.0685 | -2.1046 |

It can be seen from Table 7 and Table 8 that the government implicit guarantee has a significant impact on the rating of municipal investment bonds in both the eastern and central regions, but the effect is greater in the central and western regions than in the eastern regions.

The reason behind this may be that the eastern region is more developed, with stronger liquidity of funds and easier financing for municipal investment enterprises. The default risk of municipal investment bonds is relatively low, and the impact of implicit guarantees from local governments on credit ratings of these bonds is minimal.Compared to the eastern region, financing for municipal investment enterprises in the central and western regions is relatively difficult, and ultimately the repayment of municipal investment bonds may still rely on the government. Therefore, the implicit guarantee from local governments has a greater impact on credit ratings of municipal investment bonds.

## 5 conclusion

The article first applies text analysis methods to explore relevant policy documents related to municipal investment bonds, and calculates the implicit guarantee strength of municipal investment bonds based on the PMC index model. Then, using an ordered Logit model, it further analyzes the impact of implicit guarantee strength on credit evaluation of municipal investment bonds.Research has found that the implicit guarantee provided by the government significantly affects the credit ratings of local government financing vehicle bonds. Indeed, the government's implicit guarantee for these bonds does help improve credit ratings assigned by rating agencies.And, the role of implicit government guarantees in improving credit ratings is more evident in underdeveloped areas in the western region.The reason may lie in the fact that the eastern region has a more developed economy, making it relatively easier for municipal investment enterprises to obtain financing and reducing the impact of implicit government guarantees. On the other hand, municipal investment enterprises in central and western regions face difficulties in obtaining financing, relying on government guarantees for debt repayment, thus resulting in a more significant impact.

Based on the research findings of this article, in order to mitigate the impact of implicit government guarantees on credit ratings of municipal investment bonds, it is suggested to improve relevant mechanisms from aspects such as government, municipal investment platforms, and rating agencies.Firstly, accelerate the pace of substantial separation between implicit government guarantees and financing platform debts to avoid government intervention in defaulting behaviors of local government financing vehicles.Allowing platform bonds with poor profitability and low contribution to regional economic development to default aims to change the market investment logic, making investors pay more attention to the risks of bonds.Secondly, it is necessary to enhance the profitability of municipal investment platforms and reduce their reliance on government subsidies and implicit guarantees.By optimizing business models, improving operational efficiency, and enhancing risk management capabilities, we aim to enhance the ability to generate sustainable cash flow.Ensure that the municipal investment platform follows market principles in all aspects of debt issuance, investment, and debt repayment to reduce default risk.Reform the government-dependent local government financing platforms to be market-oriented and clarify their status as market entities.By introducing social capital and implementing mixed-ownership reforms, we aim to promote the development of municipal investment platforms towards marketization and specialization.Lastly, it is necessary to strengthen the supervision of rating agencies to ensure their rating activities are compliant and legal.At the same time, it is encouraged to strengthen self-discipline and improve the overall quality and service level of the industry for credit rating agencies.

# References

[1] Ji Y. and Cao H.M. The Development History and Credit Rating Method of "municipal Investment Bonds". Macroeconomic Management, 2014, (6): 56-59.

[2] Ma W.T. and Ma C.Y. Government Guarantee Intervention, Stabilization Growth Constraints, and the Expansion Trap of Local Government Debt . Economic Research, 2018, 53(05): 72-87.

[3] Luo R.H and Liu J.J. Are Implicit Government Guarantees Effective for Local Governments? - An Empirical Test Based on the Issue Pricing of Municipal Investment Bonds . Financial Research, 2016, (4): 83-98.

[4] Ma W.T. and Zhang P. Government Implicit Guarantee, Marketization Process and Efficiency of Credit Allocation. Financial Research, 2021, (8): 91-106.

[5] Zhong N.H, Chen S.S and Ma H.X, et al. Evolution of Local Government Financing Platform Debt Risks - Measuring Expectations of "Implicit Guarantee" [J]. China Industrial Economy, 2021, (4): 5-23.

[6] Mao R., Liu N.N and Liu R. The Impact of Local Debt Financing on the Effectiveness of Government Investment: A Study. World Economy, 2018, 41(10): 51-74

[7] Wang L. and Chen S.Y. Government Implicit Guarantee, Debt Default and Interest Rate Determination. Financial Research, 2015, (09): 66-81

[8] Yuan L.P and XiaoY. The Impact of the New Policy on Managing Local Government Debt on the Credit Risk of municipal Investment Bonds . Financial Development Research, 2017, (06): 3-9.

[9] Zhu X.Q., Chen Z., and Shi Z., et al. The Hedging Effect and Spillover Effect of Contagion of Default Risk: Perspective of Implicit Guarantee Expectation. Economic Research, 2022, 57(11): 174-190.

[10] Zhong H.Y, Zhong N,H, and Zhu X.N. Is the Guarantee of Urban Infrastructure Bonds Trustworthy? Evidence from Bond Ratings and Issue Pricing. Financial Research, 2016, (4): 66-82.

[11] Chen Y.F. Can Implicit Government Guarantees Effectively Lower the Credit Spread of Municipal Bonds? - Based on the Regression Results of Bonds with Different Ratings. Guangxi Quality Supervision and Inspection Guide, 2020, (09): 98-99.

[12] Wang B.S, Lü Y.Q and Ye Y.X. Government Implicit Guarantee Risk Pricing: An Exploration Based on China's Bond Trading Market. Economic Research, 2016, 51(10): 155-167.

[13] Ma G.J and Sun S. Credit Rating Indicators System for " municipal Investment Bonds" Based on Risk Specificity. Development Research, 2011, (6): 78-81.